\pdfoutput=1

\documentclass[]{spie}

\usepackage[]{graphicx}
\usepackage{aasmacros}
\usepackage{rotating}

\def\deg{\ensuremath{^\circ}}
\newcommand{\prog}[1]{\emph{#1}}
\def\fcp{\prog{fcp}}

\newcommand{\code}[1]{{\tt #1}}
\def\algman{\code{algMan}}
\def\udps{\code{UDPS\_Listener}}

\title{Software systems for operation, control, 
and monitoring of the EBEX instrument}

\author{
Michael Milligan,\supit{a} 
Peter Ade,\supit{b} 
Fran\c cois Aubin,\supit{c} 
Carlo Baccigalupi,\supit{d} 
Chaoyun Bao,\supit{a} 
Julian Borrill,\supit{e} 
Christopher Cantalupo,\supit{e} 
Daniel Chapman,\supit{f} 
Joy Didier,\supit{f} 
Matt Dobbs,\supit{c} 
Will Grainger,\supit{b} 
Shaul Hanany,\supit{a} 
Seth Hillbrand,\supit{f} 
Johannes Hubmayr,\supit{g} 
Peter Hyland,\supit{c} 
Andrew Jaffe,\supit{h} 
Bradley Johnson,\supit{i} 
Theodore Kisner,\supit{e} 
Jeff Klein,\supit{a} 
Andrei Korotkov,\supit{j} 
Sam Leach,\supit{d} 
Adrian Lee,\supit{i} 
Lorne Levinson,\supit{k} 
Michele Limon,\supit{f} 
Kevin MacDermid,\supit{c} 
Tomotake Matsumura,\supit{l} 
Amber Miller,\supit{f} 
Enzo Pascale,\supit{b} 
Daniel Polsgrove,\supit{a} 
Nicolas Ponthieu,\supit{m} 
Kate Raach,\supit{a} 
Britt Reichborn-Kjennerud,\supit{f} 
Ilan Sagiv,\supit{a} 
Huan Tran,\supit{i} 
Gregory S. Tucker,\supit{j} 
Yury Vinokurov,\supit{j} 
Amit Yadav,\supit{n} 
Matias Zaldarriaga,\supit{n} 
Kyle Zilic\supit{a}
\skiplinehalf
\supit{a}University of Minnesota School of Physics and Astronomy, Minneapolis, MN 55455; \\
\supit{b}Cardiff University, Cardiff, CF24 3AA, United Kingdom; \\
\supit{c}McGill University, Montr\'eal, Quebec, H3A 2T8, Canada; \\
\supit{d}Scuola Internazionale Superiore di Studi Avanzati, Trieste 34151, Italy; \\
\supit{e}Lawrence Berkeley National Laboratory, Berkeley, CA 94720; \\
\supit{f}Columbia University, New York, NY 10027; \\
\supit{g}National Institute of Standards and Technology, Boulder CO 80303; \\
\supit{h}Imperial College, London, SW7 2AZ, England, United Kingdom; \\
\supit{i}University of California, Berkeley, Berkeley, CA 94720; \\
\supit{j}Brown University, Providence, RI 02912; \\
\supit{k}Weizmann Institute of Science,  Rehovot 76100, Israel; \\
\supit{l}California Institute of Technology, Pasadena, CA 91125; \\
\supit{m}Institut d'Astrophysique Spatiale, Universite Paris-Sud, Orsay, 91405, France; \\
\supit{n}Institute for Advanced Study, Princeton, NJ 08540
}

\begin{document}
\maketitle

\begin{abstract}

We present the hardware and software systems implementing autonomous
operation, distributed real-time monitoring, and control for the EBEX
instrument.  EBEX is a NASA-funded balloon-borne microwave polarimeter
designed for a 14 day Antarctic flight that circumnavigates the pole.  

To meet its science goals the EBEX instrument autonomously executes
several tasks in parallel: it collects attitude data and maintains
pointing control in order to adhere to an observing schedule; tunes
and operates up to 1920 TES bolometers and 120 SQUID amplifiers
controlled by as many as 30 embedded computers; coordinates and
dispatches jobs across an onboard computer network to manage this
detector readout system; logs over 3~GiB/hour of science and
housekeeping data to an onboard disk storage array; responds to a
variety of commands and exogenous events; and downlinks multiple
heterogeneous data streams representing a selected subset of the total
logged data.  Most of the systems implementing these functions have
been tested during a recent engineering flight of the payload, and
have proven to meet the target requirements.

The EBEX ground segment couples uplink and downlink hardware to a
client-server software stack, enabling real-time monitoring and
command responsibility to be distributed across the public internet or
other standard computer networks.  Using the emerging dirfile standard
as a uniform intermediate data format, a variety of front end programs
provide access to different components and views of the downlinked
data products.  This distributed architecture was demonstrated
operating across multiple widely dispersed sites prior to and during
the EBEX engineering flight.

\end{abstract}


\section{INTRODUCTION}
\label{sec:intro}

\subsection{Science goals}
\label{ssec:science}

The E and B EXperiment (EBEX) is balloon-borne microwave polarimeter
designed to study the polarization of the cosmic microwave background
(CMB) and the foreground emission of thermal dust in our
galaxy. \cite{oxley04-alt,grainger08-alt} These measurements will:
detect or constrain the primordial B-mode polarization of the CMB, a
predicted signature of gravity waves produced by cosmic inflation;
\cite{seljak97,Hu98} characterize the polarized foreground dust
emission, which is a necessary step in determining the CMB B-mode
signal\cite{brandt94,baccigalupi03}; and measure the predicted
effect of gravitational lensing on the
CMB\cite{zaldarriaga98}.  The science goals of EBEX are described
more fully in other publications
\cite{oxley04-alt,grainger08-alt,britt10:spie}.

\subsection{Instrument description}
\label{ssec:instrument}

The EBEX instrument consists of a 1.5 meter clear aperture
Gregorian-type telescope that feeds a cryogenic receiver, all of which
are mounted on the inner frame of the EBEX gondola.  Pointing control
is maintained by driving the inner frame in elevation, while a pivot
and reaction wheel turn the outer frame azimuthally relative to the balloon flight
line.  Attitude sensors including a sun sensor, star cameras,
differential GPS, gyroscopes, magnetometer, and clinometers are
mounted as appropriate on the inner and outer frames.  The flight
computers, Attitude Control System (ACS) crate, and disk storage
pressure vessels are mounted on the outer frame.  Inside the cryostat
reimaging optics focus the input radiation onto two focal planes each
carrying up to 960 transition edge sensor (TES) bolometers, up to 1920
total bolometers.  A polarimetric system, consisting of a half wave
plate (HWP)\cite{hananyAHWP05} spinning on a superconducting magnetic
bearing\cite{hull:smb05} and a wire grid, modulates polarization
information into the phase and amplitude of the component of the
radiation intensity at the focal plane corresponding to four times the
HWP rotation frequency\cite{johnson04-phd}.  The TES are read out
through SQUID amplifiers via a frequency domain multiplexing scheme
that connects up to 16 TES to each SQUID.  The SQUIDs in turn are
connected in groups of four to digital frequency-domain multiplexing
readout (DfMux) boards\cite{dobbs08}.  The design of the EBEX
instrument is detailed elsewhere
\cite{grainger08-alt,britt10:spie,sagivmg12}, and the
bolometer readout system is described in Hubmayr et
al\cite{hubmayr08-alt} and Aubin et al\cite{franky10:spie}.
EBEX completed a 13 hour engineering flight
from Ft. Sumner, New Mexico in June 2009.
In this paper we describe the software and
data flow architecture that make up the EBEX control and monitoring
systems.

\subsection{Computing and system control overview}
\label{ssec:computing}

In order to meet the science goals, EBEX autonomously executes several
tasks in parallel.

The instrument maintains real-time pointing control to better than the $0.5\deg$
requirement and logs sufficient data from the pointing sensors to allow
post-flight pointing reconstruction to better than the 9$^{\prime\prime}$
requirement.  The
pointing system can realize several predefined instrument scan modes,
as well as drift, slew, and coordinate tracking motions.  The two
redundant flight computers (see Sec.~\ref{ssec:fcp}) execute all
pointing actions synchronously, with a watchdog card selecting
the less-recently rebooted computer to control the instrument.  The
pointing system is discussed in detail by
Reichborn-Kjennerud\cite{britt10-phd}.

Both SQUIDs and TES bolometers periodically require active tuning,
such as during cycling of the sub-Kelvin adsorption
refrigerators\cite{chaseHe3}.  This instrument reads out up to 1792
of the 1920 bolometers, multiplexed through 112 SQUIDs, operated by 28 DfMux
boards.  These setup and tuning operations are managed over the
gondola Ethernet network by the flight computers, as discussed in
Sec.~\ref{ssec:das}.

Bolometers are read out at 190.73 Hz 16-bit samples.  Depending on the
multiplexing level each DfMux board reads out between 32 and 64
bolometers, producing a data stream of between 21 and 42
kilobytes/s, or 2.1 to 4.2 gigabytes per hour for the full complement
of 28 boards.  The ACS generates an additional data stream of
approximately 20 KB/s (70 megabytes per hour), and the angular
encoders on the rotating HWP produce a combined 21 KB/s (75 MB/h).
This output data is transferred over the ethernet network to the
flight computer and logged to disk.  Consequently for a 14 day flight
the onboard disk array must provide over 1.5 terabytes total storage
per redundant copy written.  The storage system is discussed in
Sec.~\ref{ssec:aoe}.

In addition to planned housekeeping operations, the possibility of
unplanned events demands that EBEX possess the ability to respond to
some exogenous contingencies, that sufficient operational data be
downlinked to enable human diagnosis of unexpected conditions, and
that the telecommanding interface be flexible enough to exercise the
full range of recovery options available in the flying hardware.  The
necessary downlink (Sec.~\ref{ssec:downlink}) is provided by a 1 Mbit/s
line-of-sight (LOS) transmitter available for roughly the first day of
flight, and a much slower TDRSS satellite relay afterwards.  The
telecommanding uplink relies on satellite relay or an HF-band LOS
transmission, and in practice is limited to less than ten 15-bit
command tokens per second.

All of the above activities can be triggered from the ground via
uplinked commands, as well as scheduled via onboard schedule files.
The scheduling system operates in local sidereal time, allowing
planned observations to account for the motion of the balloon in
longitude, which cannot be precisely known in advance.  Within the
limits of the underlying operating system, actions can be scheduled
arbitrarily far in the future.  Uplinked commands can select between
alternative stored schedules.

The communications infrastructure of the Columbia Scientific Balloon
Facility (CSBF) provides the LOS downlink signal at the launch site,
and provides connections to satellite-based telemetry and
telecommanding via the Operations Control Center in Palestine,
Texas\cite{CSBF-LDB}.  During a long duration balloon flight, many
collaboration personnel will be positioned at the launch site, while
other collaborators may be geographically dispersed.  To support this
scenario the EBEX ground segment couples uplink and downlink hardware
to a client-server software stack (see Sec.~\ref{ssec:ground} and
Fig.~\ref{fig:dataflow}).  The full high rate LOS data stream is
available at multiple client workstations at the launch site, and
portions of this data can be made available via the public internet
for remote real-time examination.  Likewise telecommanding is
forwarded over network links to the EBEX ground station and CSBF
uplink.

To meet the reliability and development time requirements of this
project we use commercially available hardware and existing software
whenever practical.  With the exception of the FPGA-based DfMux and
ACS boards, onboard computers and networking hardware are available
industrial embedded models which we have qualified in thermal and
vacuum conditions approximating balloon flight.  The ACS, many aspects
of the gondola and pointing system design, and several components of
the software chain described here originate with the BLAST
project\cite{pascale08}, and are described by Wiebe\cite{wiebe09-phd}.
The housekeeping system makes extensive use of embedded monitoring
boards\cite{elmb-lhc03} originally developed for the ATLAS
experiment at CERN\cite{atlas-control04,elmb-lhc01}.

\section{SYSTEMS}
\label{sec:systems}

\begin{figure}
\begin{centering}
\includegraphics[bb=0.866in 0.625in 10.95in 7.42in,clip,width=6in]{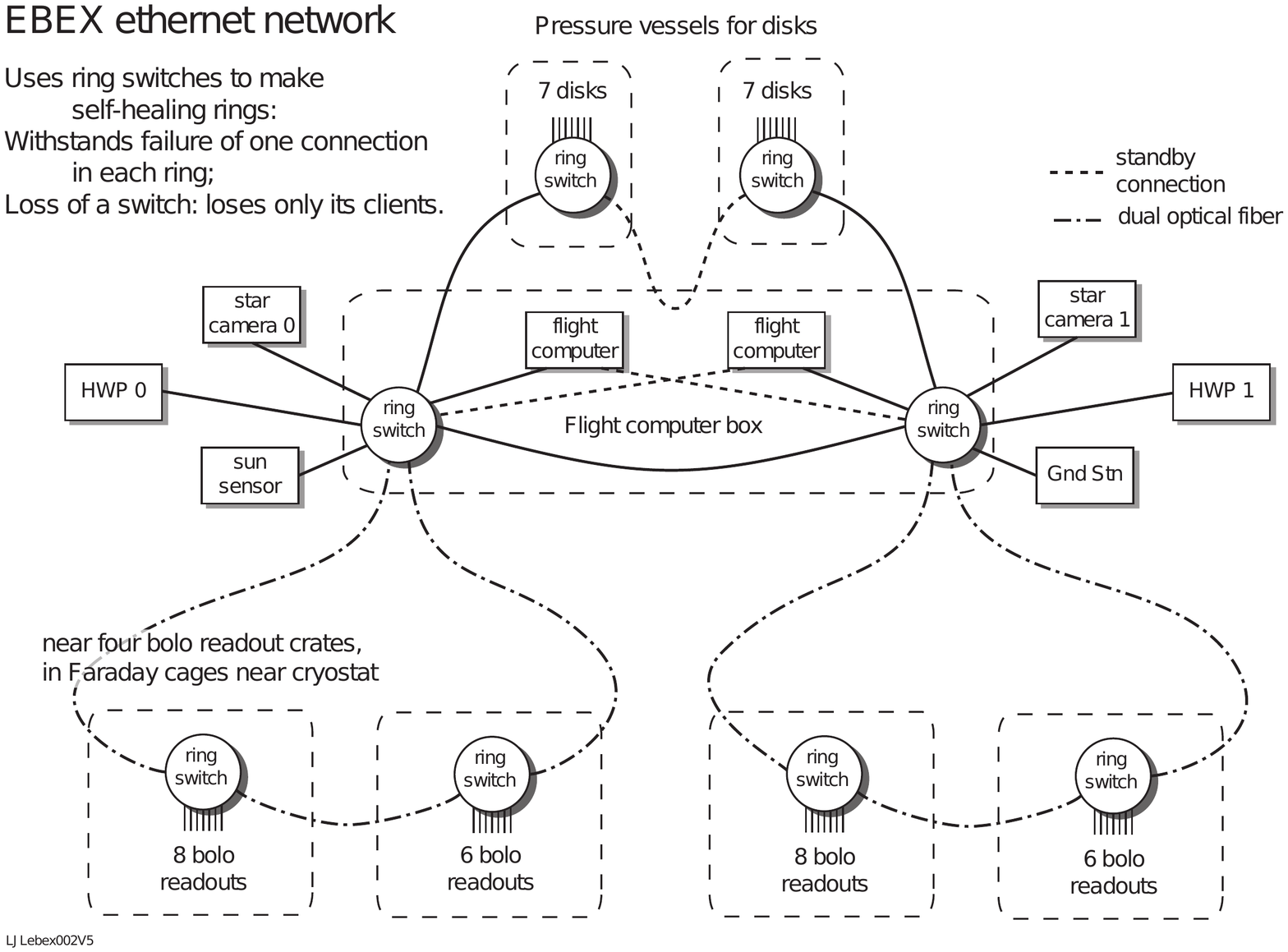}
\caption{\label{fig:ethernet} Configuration of the EBEX gondola
  ethernet network planned for the long duration Antarctic flight.}

\par\end{centering}
\end{figure}

The EBEX gondola comprises several subsystems of networked components,
with the flight computer crate acting as the point of intersection.

An Ethernet network of industrial ring switches\cite{sixnetET}
connects the flight computers, disk storage system, bolometer readout
boards, HWP encoder readouts, sun sensor, and star camera.  This
network is shown in Fig.~\ref{fig:ethernet}.  The use of ring
switches provides resilience to network breaks or failure of a single
switch.  Optical fiber connections are used where electrical isolation
is necessary.

The GPS receiver, multiple actuators, and the CSBF support package
(which includes the commanding uplink and low rate satellite telemetry)
communicate directly with the flight computers via serial
ports.  Additional sensors and controls connect directly to hardware
in the ACS crate.  The ACS communicates with the flight computers via
a custom bidirectional bus termed the ``E-bus.''\cite{wiebe09-phd,
  britt10-phd}

Housekeeping monitoring and control is handled by custom boards
equipped with embedded monitoring boards\cite{elmb-lhc01}, which are connected by a
Controller Area Network\cite{iso11898:canbus} bus (CANbus).  The
flight computers communicate with this network via Kvaser USB-CANbus
adapters\cite{kvaserleaf}.

Because the housekeeping system, ACS, and bolometer readouts are
asynchronous, all systems embed in their data streams a common
timestamp using EBEX ``ticks'' which is recorded for post-flight
alignment.  The systems maintain a relative synchronization of $\sim
10$~$\mu$s by resynchronizing to an onboard precision clock every 
164~ms.  The time servers broadcast synchronization messages onto the
CANbus, and distribute timing data to the DfMux boards and ACS via an
RS-485 serial link that does not connect to the flight computers.  The
housekeeping and timing subsystems are described in Sagiv et
al\cite{sagivmg12}.

\subsection{Flight control program -- \fcp}
\label{ssec:fcp}

\begin{sidewaysfigure}[p]
\begin{centering}
\includegraphics[width=8.5in]{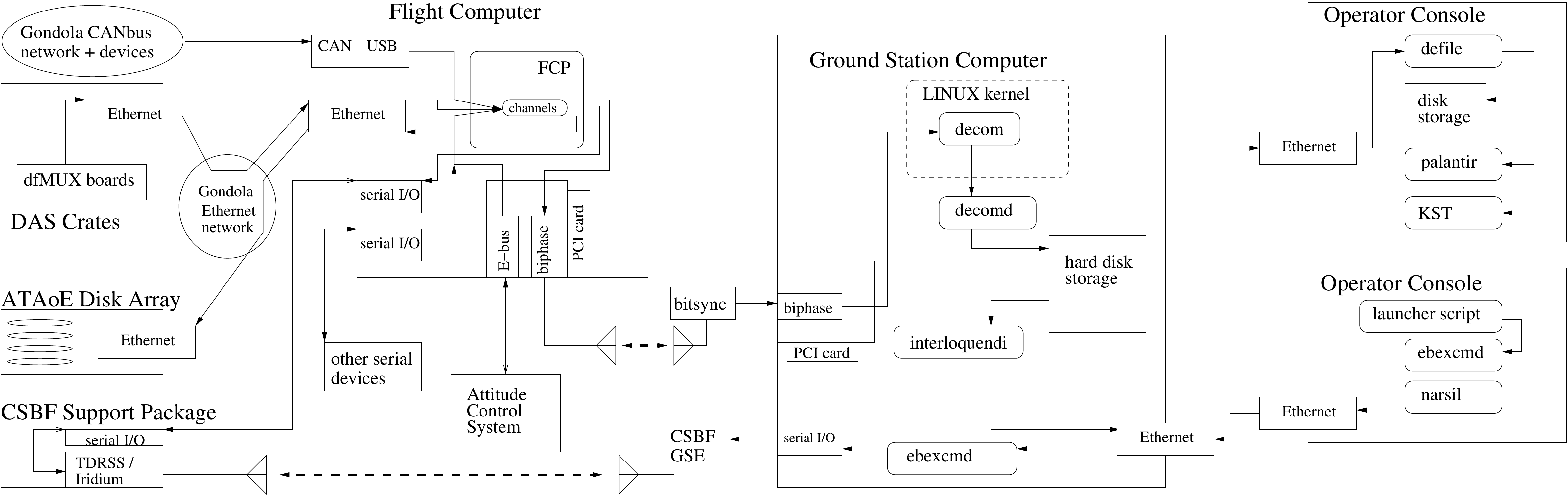}
\caption{\label{fig:dataflow} Schematic diagram of command and data
  flows in the EBEX flight and ground systems.  Square corner boxes
  represent physical components, and rounded boxes generally represent
  software modules.  The left side of the figure comprises flight
  systems, including the flight computer running \fcp{}
  (Sec.~\ref{ssec:fcp}) connected to DfMux boards in the Data
  Acquisition System (Sec.~\ref{ssec:das}) and the disk storage system
  (Sec.~\ref{ssec:aoe}).  The center of the figure represents the
  ground station, containing the interface to CSBF downlink equipment
  (\prog{biphase}, \prog{decom}, \prog{decomd},
  Sec.~\ref{ssec:downlink}), and the server portion of the data
  distribution software chain (\prog{interloquendi}).  Sample operator
  console configurations (Sec.~\ref{ssec:ground}) are shown on the
  right.  A data monitoring terminal at top illustrates the client
  (\prog{defile}) and display (\prog{KST}, \prog{palantir}) portions
  of the data distribution chain.  Below, a commanding station
  illustrates the command uplink chain via \prog{narsil} and
  \prog{ebexcmd}.  The heavy dashed lines represent radio
  communications between the gondola on the left and ground on the
  right.  In the interest of space data paths for satellite downlinks
  are omitted.
}

\par\end{centering}
\end{sidewaysfigure}

The flight computer crate contains two Ampro single board
computers\cite{amproMB}, each configured with a 1.0 GHz Celeron
processor, 256 MiB RAM, and a 1 GB solid state flash disk module.  The
module stores the computer operating system, currently Debian
GNU/Linux 4.0\cite{debianetch} with Linux kernel 2.6.18 and additional
modular drivers for the ACS E-bus and USB-CANbus adapter.  The flight
control program \fcp{} resides on the flash module as well, which the
operating system is configured to run immediately after the computer
boots.

\fcp{} is a derivative of the BLAST experiment's
\prog{mcp}\cite{wiebe09-phd}, and preserves its overall architecture
as a monolithic program running multiple concurrent, event-driven
threads, with a main loop handling pointing, frame generation, and
data logging clocked to the E-bus.  We have added code modules
implementing control and readout of the DfMux boards, housekeeping via
the CANbus, storage to the networked disk storage array, and the
downlink scheme discussed below.  Other modules have been modified as
needed.

Flight computer redundancy is implemented via a watchdog card
connected to the IEEE 1284 parallel port of each computer.  In nominal
operation the \fcp{} \code{WatchDog} thread toggles a pin on the
parallel port at 25 Hz.  If this action ceases for more than a
configurable length of time, a fault is inferred.  The watchdog card
will power cycle the faulty computer and switch control to the other
computer.  Besides crashes in the software or hardware of the flight
computer, \fcp{} can programmatically trigger this sequence of events
by terminating the \code{WatchDog} thread in response to certain error
conditions.  The identity of the computer in control is communicated
to both flight computers via the E-bus, and recorded as the
\code{incharge} variable.  During the North American engineering flight
dataset the value of this variable changes only once, at 8:19 UTC, due
to an intentional pre-flight reboot of the then-active flight
computer.  This indicates that there were no such reboots of the
in-charge flight computer between launch at 14:01 UTC and termination
after 03:18 UTC.

\subsection{Distributed networked bolometer readout architecture}
\label{ssec:das}

Each DfMux readout board combines analog signal processing hardware
with an FPGA implementing digital signal processing modules and a soft
CPU running an embedded Linux distribution.  The DfMux hardware is
described in detail by Dobbs et al\cite{dobbs08}.  Operations
comprising the setup, tuning, and maintenance of the detectors and
readout system are controlled by the flight computer via requests over
the Ethernet network, and readout data are returned over the same
network.

Low level operations are exposed via small programs in the DfMux
firmware implementing the Common Gateway Interface\cite{rfc3875}.
More complex algorithms are invoked as jobs through an interface
called ``Algorithm Manager,'' which passes data using JavaScript
Object Notation\cite{rfc4627}.  On each DfMux board a program,
implemented by code in a subset of the Python language\cite{python},
listens on a network port for requests to start, stop, or collect the
output of jobs.  Because of memory and CPU constraints in the embedded
environment, no more than two jobs may run at a time on each board.  In
\fcp{} the \algman{} module maintains queues of pending and running
jobs and attempts to run all requested jobs as soon as possible, while
ensuring that on a per-board basis all jobs are run in the order
requested.  To the rest of \fcp, \algman{} exposes routines to trigger
algorithm requests to a single board.  It also provides a higher level
interface based on stored parameter files.  In these files sets of
algorithm parameters are defined on a per-SQUID basis.  After
commanding \fcp{} to parse one of the stored files, \algman{} will
respond to these high-level commands by dispatching algorithm requests
for the corresponding operation for each SQUID defined in the
parameter file.

Regardless of the method of invocation, requested operations will
produce output strings in the JavaScript Object Notation format which are
returned to \algman.  These strings, generically termed ``algorithm
results,'' are logged to disk and added to the file downlink system
queue.

DfMux boards output data samples by broadcasting User Datagram
Protocol\cite{rfc768} packets to a
multicast address over the Ethernet network.  Each packet is 1428
bytes and consists of a header and 13 frame structures.  In the case
of the bolometer readout boards in the configuration flown in the 2009
engineering flight, with 8 bolometer per SQUID multiplexing
(32 total bolometer channels per
board) these frames contain a timestamp and one 16-bit sample for each
of the 32 channels recorded at the corresponding time.  For a 190.73
Hz sample rate each board broadcasts packets at 14.67 Hz.  Within each
bolometer readout crate, the DfMux boards are synchronized to a common
25~MHz oscillator so that the bolometers for all boards in the crate
are sampled at the same time.

\begin{figure}
\begin{centering}
\includegraphics[width=6in]{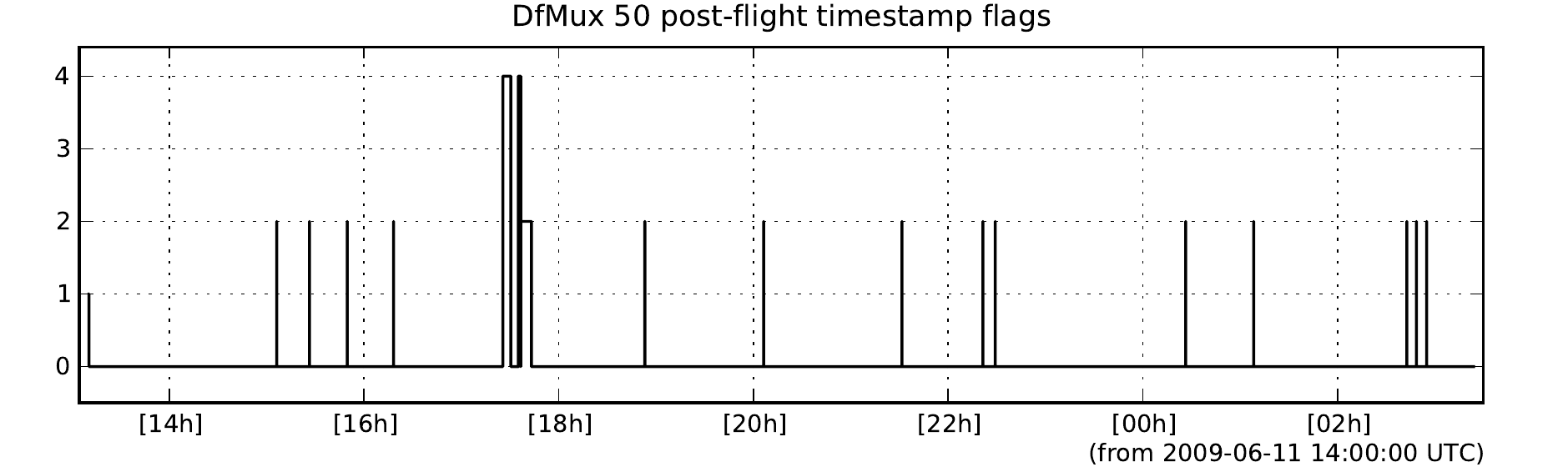} \par\end{centering}

\caption{\label{fig:syncflags}Synchronization flags for a typical
  bolometer readout board during the 2009 flight.  The flag values
  indicate: 0 -- sample present and synchronized; 1 -- padding at
  ends; 2 -- missing data; 4 -- wrong sample rate.  The anomalous
  behavior around 17:35 UTC corresponds to a commanded reboot
  of the DfMux boards.  Most of the
  isolated spikes to state 2 indicate single packets missing from the
  recorded data stream, 20 in total for this board.  Otherwise for
  this board data samples were logged for the entire flight, and those
  samples were synchronized to the common oscillator.
}

\end{figure}

In \fcp{} the \udps{} packet reader thread listens on the multicast
address.  Each packet is inspected to determine its origin, and the
\code{pdump} module writes it to disk in a packet dump (\code{.pdump})
file corresponding to the originating board.  The \code{.pdump} files
are rotated every 15 minutes to limit the maximum file size produced.
Fig.~\ref{fig:syncflags} demonstrates the performance of this readout
system for a typical readout board.  Excluding a brief period around
17:35 UTC when the boards were commanded to reboot during a SQUID
tuning procedure, no board is missing more than
65 packets from the logged packet data, for a loss rate of $<0.01$\%.
Testing on the ground shows that under simulated  load equivalent to the full
planned complement of 28 boards, loss rates remain similarly low.
11 of the 12 bolometer readout boards were synchronized to the common
oscillators in their respective crates for the entire flight.
The twelfth board was left unsynchronized due to a misconfigured startup script.

Two DfMux boards are also used to read the optical angular encoder on
the HWP.  They each sample a single channel at 3.052 KHz.  Each HWP
encoder packet contains 416 samples, and thus each board broadcasts
packets at 7.34 Hz.  The structure of the bolometer readout packets is
reused for the HWP encoder readout, so the same code processes both
types of packet stream.

The code defining the packet format is written in portable C that is
compiled into the packet streamer program onboard the DfMux CPU,
\udps, and the standalone \code{parser} program used to extract data
from packet streams and saved dumps.

\subsection{ATAoE onboard storage}
\label{ssec:aoe}

EBEX will fly with over 3 terabytes of hard disk storage.  This
allows the flight computers to write two redundant copies of all data
produced in flight to separate disks.
We use the ATA over Ethernet (ATAoE) protocol\cite{ataoe} in order to
implement the onboard disk storage array.  Ethernet has several
attractive features.  It provides a many-to-many topology so that
redundant disks can be provided without foreknowledge of which flight
computer will need one.  It is physically straightforward to route
signals from the flight computer crate in vacuum into the pressure
vessels holding hard disks.  Finally, Ethernet is already in use
onboard so it avoids adding an additional networking technology.
Drivers for the ATAoE protocol are a standard part of the Linux kernel.

As shown in Fig.~\ref{fig:ethernet}, the disk drives are divided
between two pressure vessels.  Each vessel contains a ring switch, a passive backplane
for power and signal distribution, and up to seven 2.5'' laptop disk
drives mounted on ATAoE blades\cite{coraid}.  Each blade is connected
independently to the ethernet ring switch.  In \fcp{} the
\code{EBEX\_AOE} module abstracts detection, setup and low-level
management of the array.  Disk usage is flagged in non-volatile memory
present on each blade to ensure that the two flight computers do not
attempt to simultaneously mount the same disk.  This module will only
present as available disks which are not already in use and which have
sufficient free space remaining.  The \code{aoeMan} module adds an
additional layer of abstraction, allowing \fcp{} code to request file
operations without any detailed knowledge of the disk array.  It
mounts disks as needed to supply the requested free space, and
translates filenames to correspond with the correct mount points in
the filesystem namespace.

\subsection{Downlink and data logging}
\label{ssec:downlink}

\fcp{} produces a 1 Mbit/s biphase encoded output data stream,
suitable for transmission over the CSBF-provided line-of-sight downlink.  This
stream combines all output channels of the ACS and housekeeping
systems, packet data streams from five selectable DfMux boards, and a
file downlink system called \code{filepig}, used to retrieve algorithm
results, diagnostic logs, and other irregularly formatted data.  At
the launch site the EBEX ground station uses a commercial bit
synchronizer, custom decommutator card, and the \prog{decomd}
software to decode and store this data stream to disk.

\begin{figure}
\begin{centering}
\includegraphics[width=6.5in]{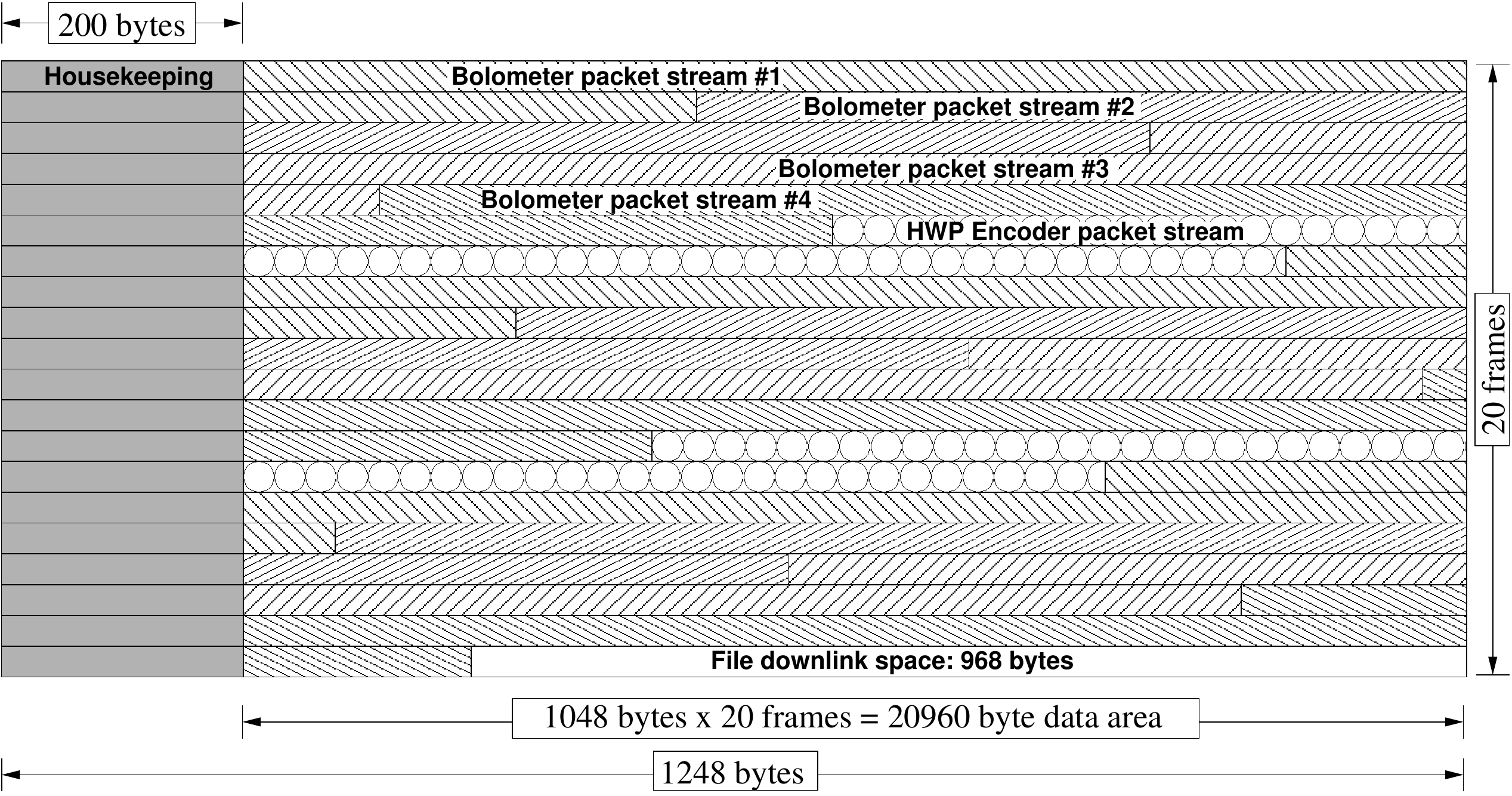} 
\caption{\label{fig:superframe} Schematic of the line-of-sight
  downlink superframe discussed in Sec. \ref{ssec:downlink}.  This
  structure is repeated at 5~Hz over the 1 Mbit/s transmitter.
  The horizontal rows indicate the 20 individual 1248 byte frames,
  transmitted at 100~Hz.  Each frame starts with 200 bytes of header
  and housekeeping channel data.  The remaining 1048 bytes in each
  frame is aggregated across the superframe to form a 20960 byte data
  area.  14 slots of 1428 bytes each are allotted for DfMux packets
  and are grouped into five logical streams (denoted here by matching
  hatch patterns), accomodating the complete data output of four
  bolometer readout boards and one HWP encoder readout board.  The
  final 968 bytes of the superframe is used by the \code{filepig} file
  downlink system.
}
\par\end{centering}
\end{figure}

As detailed in Fig.~\ref{fig:superframe}, the downlink stream is
composed of 1248 byte frames generated at
100~Hz.  These are grouped into superframes of 20 frames.  Each frame
begins with a sync word and counters, followed by channel data.  Each
2-byte word of channel data can either contain samples of a ``fast
channel'' at 100~Hz, or have 20 ``slow channels'' multiplexed over the
superframe at 5~Hz.  In the 2009 engineering flight, this channel data
totalled 194 bytes per frame, encoding 59 fast channels and 480 slow
channels.  This channel data is also logged to disk onboard the
gondola.

The remaining space in each frame (1048 bytes, after overhead, for the
2009 flight configuration) is aggregated across the superframe and
used to transfer DfMux readout packets and \code{filepig} data blocks.
In \fcp{} this format is defined by the ``Biphase marshaler''
module, which accepts data from \udps{} and \code{filepig} and
assembles the superframe data area.  Every 200~ms the \fcp{} downlink
code queries the marshaler for an assembled data area to incorporate
into the transmitted frames.

The marshaler uses fixed slots in the superframe to provision a
deterministic bandwidth to each downlinked data stream, and to ensure
that if one frame is lost or corrupted, data in the surrounding frames
can still be correctly reassembled.  \udps, described above, passes
whole packets, and thus requires 1428-byte slots.  In 200 ms a
bolometer readout board produces on average 2.93 packets, and a HWP
encoder board produces 1.47.  Thus a group of three slots for
bolometer readout or two slots for encoder readout yields a stream
with adequate capacity to downlink the entire packet data output of a
DfMux board.  With 14 slots, streams are defined to downlink the output
of four bolometer readout boards and one HWP encoder board.  Uplinked
commands select which five boards out of the total complement are
allotted a downlink stream.

\code{filepig}, so named because it allows files to ``piggyback'' on a
frame-based protocol, claims the odd-sized chunk of space at the end
of the data area after packet streams have been allocated.  It exposes
an interface by which \fcp{} code may queue the filenames of data
objects already written to disk.  Files are broken into chunks
together with minimal header and error detection data and downlinked.
Support exists, presently unused, to plug in transformations for more
robust error correction or data compression, and to resend corrupted
data in response to uplinked commands.  For the engineering flight 968 bytes
per superframe were left for the \code{filepig} data chunk, providing
about 4.2 KB/s file downlink bandwidth.  Over the 13 hour flight
10898 files totalling 61 MB were retrieved.

\subsection{Ground tools and architecture}
\label{ssec:ground}

The BLAST telemetry chain\cite{wiebe09-phd} is employed largely
unmodified on the ground.  As illustrated in Fig.~\ref{fig:dataflow},
the biphase encoded bitstream is converted back into data frames in
the Ground Station computer and logged to disk.  The
\prog{interloquendi} server permits clients to fetch streams of frames
remotely via TCP/IP connections. \prog{defile} then decodes the
channel data in these frames into dirfile\cite{dirfile}-format data
files.  Front end programs such as \prog{palantir} and
\prog{KST}\cite{kst-kde} allow real-time display of the streamed
channels.

To this EBEX adds support in the frame handling code for the
superframe data area, and support in \prog{defile} for extracting
packet streams and downlinked files from those frames.  These
additional data products are written alongside the channel-based data
on each connecting client workstation.  Scripts employing the
\prog{parser} program automate the production of bolometer and HWP
encoder time streams in dirfile format from extracted \code{.pdump}
files.

Time streams can be displayed in real-time using either \prog{KST} or
Python tools that understand the dirfile format.  The EBEX Alignment
Tools is a suite of programs for further processing these streams,
including interpolation and alignment to a common sample rate and
timing, decoding the HWP angular encoder signal to HWP position, and
template-based removal of the HWP rotation signal from bolometer
timestreams.

We have also written a Python/TK front end to the Algorithm Manager
system.  By monitoring the names of the files downlinked through
\code{filepig}, it is possible to select those corresponding to
algorithm result strings.  Parsing these files permits display on a
board-by-board basis, in close to real time, of the job execution
activity occurring in the readout system DfMux boards.  A dashboard
interface presents selected information from each board using labels
and color coding, and the user can select individual boards or jobs
for more detailed display.  This front end provides immediate visual
feedback on complex operations, such as detector system tuning, that
entail parallel execution of a sequence of jobs on each bolometer
readout board.

\prog{ebexcmd} accepts \fcp{} commands in textual format, which it can
either relay to a listening \prog{ebexcmd} over a network connection,
or convert to the binary representation suitable for transmission over
CSBF uplink hardware.  Commands can therefore be generated on any host
permitted to connect to the ground station, and those commands will
then be uplinked.  Commands are most commonly selected through the
\prog{narsil} front end, but are also generated by Python scripts and
may even be entered manually from a command line.

This ground infrastructure provides network transparency in both data
distribution and commanding, allowing flight operators to monitor and
control the instrument from an arbitrary number of networked
workstations.  During the 2009 integration campaign and flight, this
system routinely connected as many as ten client workstations
over the private internal network at the New Mexico launch site.  Late
in the flight line-of-sight communications were only possible from the downrange
station in Arizona, and commands were successfully relayed from the
launch site through the downrange ground station \prog{ebexcmd}.
Streaming of frame data via \prog{interloquendi} from the downrange
station to the launch site, and from the launch site to collaborators
at their home institutions, worked only intermittently due to
bandwidth constraints at the launch site.

\section{CONCLUSION}

EBEX combines a large format bolometer array, and the correspondingly
large data volume, with a complex readout system architecture.  As a
result, EBEX solves for a balloon flight environment problems in data
handling, communications, and control that are typically associated
with ground based observatories.  The required 3~terabyte in-flight
storage capacity is achieved using a high speed gondola ethernet
network and networked disk storage arrays.  The readout system is
controlled from a central flight computer using a custom distributed
job control scheduler, and it is monitored by extending a
frame-oriented telemetry system to support asynchronous packet streams
and event-driven downlink of arbitrary data in files.  On the ground,
a networked real-time data distribution and command relay architecture
allows shared monitoring and control of the instrument.

\section*{ACKNOWLEDGMENTS}

EBEX is supported by NASA through grants number
NNG04GC03G, NNG05GE62G, NNX08AG40G, and
NNX07AP36H.  Additional support comes from the National Science
Foundation through grant number AST 0705134, the French Centre national
de la recherche scientifique (CNRS), and the UK Science and Technology
Facilities Council (STFC).  This project makes use of the resources of
the Minnesota Supercomputing Institute and of the National Energy
Research Scientific Computing Center (NERSC), which is supported by
the office of Science of the U.S. Department of Energy under contract
No. DE-AC02-05CH11231. The McGill authors acknowledge funding from the
Canadian Space Agency, Natural Sciences and Engineering Research
Council, Canadian Institute for Advanced Research, Canadian Foundation
for Innovation and Canada Research Chairs program.

We thank the Columbia Scientific Balloon Facility for their energetic
support.  We gratefully acknowledge the efforts of our colleagues in
the BLAST project that produced the foundation for much of this work.

\bibliography{refs}
\bibliographystyle{spiebib}

\end{document}